\documentclass[fleqn,10pt]{wlscirep}
\usepackage[utf8]{inputenc}
\usepackage[T1]{fontenc}
\title{How initial distribution affects symmetry breaking induced by panic in ants: experiment and flee-pheromone model}

\author[1,+]{Geng Li}
\author[2,+]{Weijia Wang}
\author[1]{Jiahui Lin}
\author[1]{Zhiyang Huang}
\author[1]{Jianqiang Liang}
\author[3]{Huabo Wu}
\author[1]{Jianping Wen}
\author[2]{Zengru Di}
\author[4]{Bertrand Roehner}
\author[2,*]{Zhangang Han}
\affil[1]{School of Physics and Optical Information Sciences, Jiaying University, Meizhou, 514015, China}
\affil[2]{School of Systems Science, Beijing Normal University, Beijing, 100875, China}
\affil[3]{School of Electronic and Information Engineering, Jiaying University, Meizhou, 514015, China}
\affil[4]{Institude for Theoretical and High Energy Physics, University of Paris 6 (Pierre and Marie Curie), Paris, F-75005, France}

\affil[*]{zhan@bnu.edu.cn}

\affil[+]{these authors contributed equally to this work}

\begin{abstract}
Collective escaping is a ubiquitous phenomenon in animal groups. Symmetry breaking caused by panic escape exhibits a shared feature across species that one exit is used more than the other when agents escaping from a closed space with two symmetrically located exists. Intuitively, one exit will be used more by more individuals close to it, namely there is an asymmetric distribution initially. We used ant groups to investigate how initial distribution of colonies would influence symmetry breaking in collective escaping. Surprisingly, there was no positive correlation between symmetry breaking and the asymmetrically initial distribution, which was quite counter-intuitive. In the experiments, a “flee stage” was observed and accordingly a flee-pheromone model was introduced to depict this special behavior in the early stage of escaping. Simulation results fitted well with the experiment. Furthermore, the flee stage duration was calibrated quantitatively and the model reproduced the observation demonstrated by our previous work. This paper explicitly distinguished two stages in ant panic escaping for the first time, thus enhancing the understanding in escaping behavior of ant colonies.
\end{abstract}
\begin{document}

\flushbottom
\maketitle
%
%
\thispagestyle{empty}

\section*{Introduction}

Panic would be induced within animal groups once dangerous factors are probed, predators for instance, in natural conditions. As a direct response to panic, dramatic change in behavioral pattern of individuals can be observed, usually from stable to highly dynamic state within the whole group. Typically, individuals would escape from the region where they detect threats. For animals with high sociality, obviously behavioral patterns can been observed in panic escape. Numerous studies both in empirical and theory have been performed to investigate panic escape across species, such as ant colonies\cite{Altshuler2005,Boari2013,Parisi2015,Soria2012,Chung2017}, mouse groups\cite{Saloma11947,Saloma2015} and human crowds\cite{Helbing2000,Helbing2002,Helbing2005}. Symmetry breaking induced by panic escape, with one exit used more than the other when individuals try to escape from a room with two symmetrically located exits has aroused broad attention. This phenomenon was observed both in human evacuation in closed environment by Helbing et al\cite{Helbing2000} and different ant species, including \textit{Solenopsis invicta} Buren\cite{Li2014}, \textit{Polyrhachis dives}\cite{Chung2017}, \textit{Polyrhachis vicina Roger}\cite{Ji2018} and \textit{Camponotus japonicas}\cite{Wang2016}. 

As for collective behavior, symmetry breaking would be observed once the distribution of individuals in the environment is not uniform, but concentrates in some specific regions or directions. This special phenomenon can be widely found in animal groups with high sociality, significantly promoting survival ability of the species. As an typical example, when dealing with foraging mission, symmetry breaking in spatial distribution of the colony means decision-making among available choices and its degree reflects the allocation of labors. Specifically, sufficient studies in experiment have demonstrated that the type of food/nutrition condition of the colony\cite{Portha2002} and difference in the quality of food sources(concentration of carbohydrate or proteins)\cite{Price2016} can all affect the asymmetric distribution of ants in foraging, the allocation of labors among available food sources is determined consequently. On the other hand, these spatial asymmetries are usually the result of a self-organized process which is a shared feature of symmetry breaking induced by panic. However, the behavior of ants under panic shows quite different patterns with those in foraging. In panic, alarm pheromones will be used and specific behavior responses will emerge\cite{Wilson1971}.Specifically, alarm pheromones generally consist of low-molecular weight, highly volatile compounds that easily spread in local area, making it possible to result in macro panic escape eventually. Therefore, experiments under panic conditions are necessary to investigate the factors that can influence panic-induced symmetry breaking, namely the escaping symmetry breaking. 

It has been sufficiently discussed that the group density\cite{Li2014}, temperature\cite{Chung2017} and the width of escaping exits\cite{Ji2018} all play major roles in escaping symmetry breaking. These experimental studies were performed with the consideration that ants are evenly distributed in space (or just neglected the distribution condition\cite{Chung2017}), the initial distribution of group is thus ignored or seen as a trivial factor that barely affects the escaping symmetry breaking. Intuitively, however, it is reasonable to infer that panic-induced symmetry breaking is related to the group initial distribution when escape begins. As for a closed room with two symmetrically located exits, if there are more ants near the left exit than the right at the beginning of the escape, more ants would probably escape from the left and thus resulting in symmetry breaking; furthermore, with the increase of the initial difference in the amount of ants near two exists, the level of escaping symmetry breaking is likely to increase consequently. In fact, this kind of asymmetric distribution would be observed when groups aggregate in natural conditions, from the temporal aggregation of mosquitoes\cite{Attanasi2014,Attanasi2014a} to the schooling of herring\cite{Allee1927,Parrish1999} and flocking of starling\cite{JohnT.Emlen1952,Ballerini2008}. Especially for ants, spatial aggregation can be the initiation of some more complex collective behaviors\cite{Cole1996,Theraulaz2002,Mlot2011,Vernerey2018}. However, to what extent the initial distribution can influence the escaping symmetry breaking remains uninvestigated. 

To reveal the underlying mechanisms leading to escaping symmetry breaking, Altshuler et al proposed a Vicsek-like model based on the orientation alignment with neighbors\cite{Altshuler2005}, but the model cannot explain the density dependence of escaping symmetry breaking reported in our previous work\cite{Li2014}.This result suggests that a Vicsek-like model with an alignment rule may not be the correct model for escaping ants. Based on biological facts that ants use pheromones to communicate, rather than seeing how other individuals move, we proposed a simple yet effective alarm pheromone model. By quantitatively reproducing the first rising and then declining relationship between escaping symmetry breaking and group density\cite{Li2014}, alarm pheromone model showed its effectiveness in explaining ant panic behavior. Alarm pheromone as a possible media for ants to transfer information in panic can be validated in this degree. Therefore, we firstly used this model to give a prediction on relationship between initial distribution and escaping symmetry breaking. As a measurement of initial distribution, Initial Difference was defined, to characterize the percentage difference in the amount of ants between the right and left part of the experimental cell. Then five different Initial Difference values were simulated with alarm pheromone model. With the increase of Initial Difference, escaping symmetry breaking showed an increasing trend, which seems plausible. However, the experiment is still necessary to validate this forecast. We designed a panic escaping experiment similar with previous work\cite{Li2014}, but with initial distribution specifically considered. Surprisingly, the result of the experiment is obviously different from that of the alarm pheromone model, which is quite counter-intuitive. The alarm pheromone model fails to explain the current experiment phenomenon. 

Pheromone is defined as a substance released by an organism to the outside that causes a specific behavioral or physiological reaction in a receiving organism of the same species\cite{Nordlund1976}. For ants, this special chemical can be made in exocrine glands found throughout the ant’s body and a major medium in information transfer among individuals. Thus, pheromone trails will emerge through positive feedback and can be significantly important for mass-foraging\cite{Wilson1962}, available nest searching\cite{Cronin2012}, recruiting to battlegrounds and retreat\cite{Allee1927,HOLLDOBLER1976} and guiding the tunnel building\cite{Theraulaz1998}(\cite{Czaczkes2015} to see a review), making it possible for colonies to accomplish specific missions in a self-organization way without global information. As for panic escape, pheromone also plays an important role in a similar way\cite{Wilson1971} and has been demonstrated by alarm pheromone model\cite{Li2014}. Hence, to explain the counter-intuitive results of the current experiment, it is reasonable to construct the theoretical model based on modification of alarm pheromone model. 

To find out details potentially ignored in alarm pheromone model, we analyzed the videos recording procedures of current experiments and found a special stage at the very beginning of the panic escape. In this stage, ants would move "nearly straight" with random orientations. It seems that pheromone may not be dominant in individual’s behavior control as ant behaviors are similar to special random movements, namely with "relatively straight trajectories" but stochastic direction. The "nearly straight" is used here for trajectories of ants are arcs and broken lines in fact, however, with small tortuosity at each step. These trajectories can be seen as nearly straight based on observation. This pattern of trajectories has also been demonstrated by Angilletta et al\cite{Angilletta2007}. In their work, the fractal dimension of ant trajectories is above 1.0, meaning that ants would not move in straight strictly in panic escape. We define this stage as “flee stage”, for ants are fleeing from danger they have probed. Flee stage can be important to explain the current experimental results as the "moving nearly straight" tends to result in random distribution which largely offsets the initial asymmetry distribution. After the flee stage, we assume that ant’s behavior would be determined by pheromone and we define this stage as “pheromone stage”. A flee-pheromone model with two consequent stages was thus constructed and quantitatively reproduced the counter-intuitive experiment phenomenon. 

Our previous work\cite{Li2014} have demonstrated that with the density of the ant colony increases, panic-induced symmetry breaking would display a first increase then decrease trend. This observation can be used to calibrate the duration of flee stage to a more reasonable range, enabling flee-pheromone model a generalization ability to explain symmetry breaking in current and previous experiment. Alarm pheromone model can be a special case of flee-pheromone model when duration of the flee stage equals 0. With these efforts, we proposed a uniform model in this paper to uncover the underlying mechanism from which escaping symmetry breaking can emerge. 

\section*{Results}

\subsection*{Alarm pheromone model and panic escaping experiment}

\begin{figure}[ht]
\centering
\includegraphics[width=\linewidth]{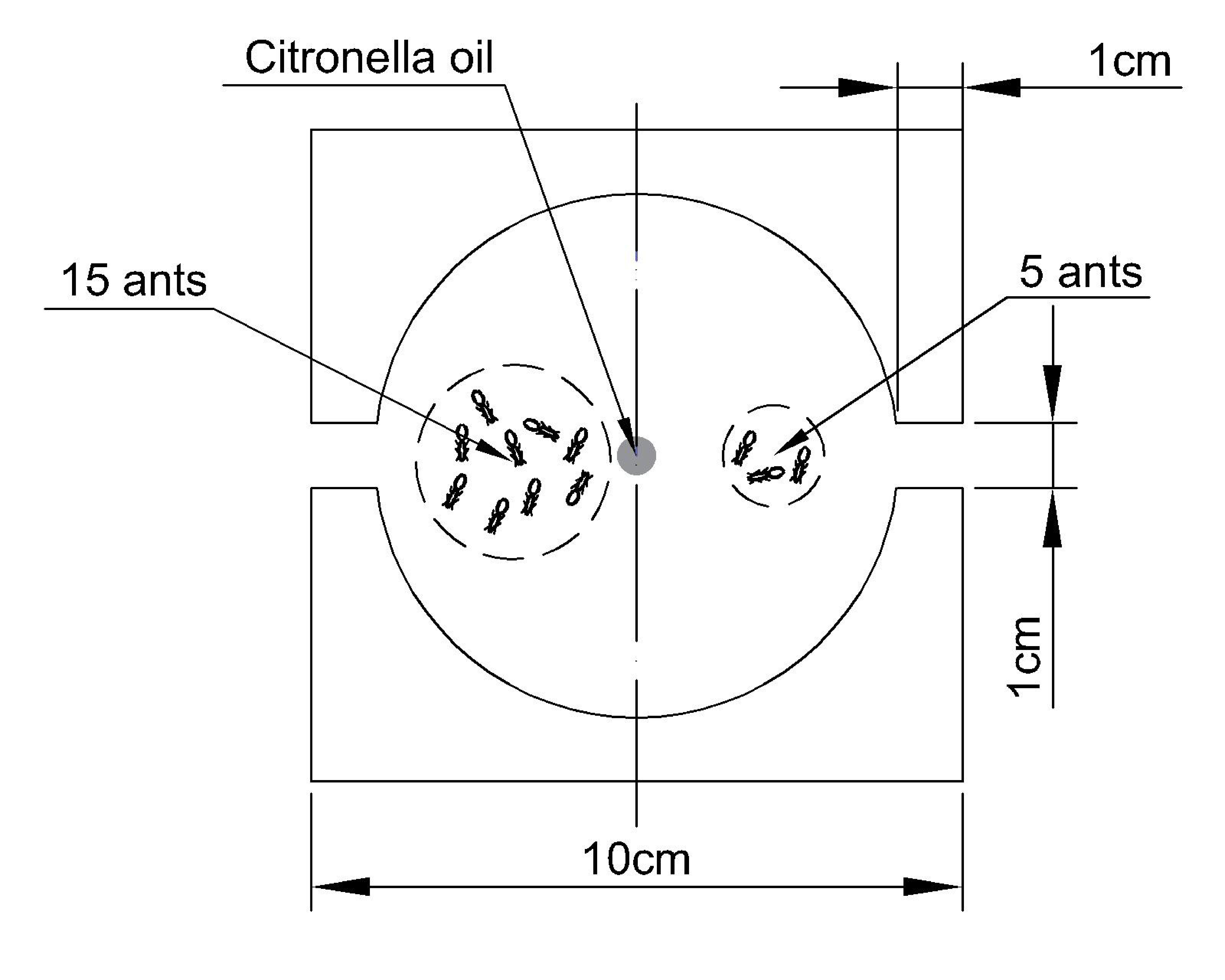}
\caption{\textbf{Schematic diagram of the experimental device for an Initial Difference 0.5.} The circular cell can be divided virtually into left and right part equally by the dotted line in center. 20 ants are initially placed in two groups according to the value of Initial Difference, 15 for the left and 5 for the right in this case(ants in the diagram are for illustration), and then a dose of 50 $\mu L$ citronella oil (a kind of repellent fluid shown as the grey area) is introduced into the center to induce panic. Ants would thus escape from the cell from the two symmetrically located exits in the left and right part of the cell, both 1 cm in width.}
\label{fig:Fig_1}
\end{figure}

Ants will be attracted towards the pheromone source at low concentration and at high concentration go into frenzied activity, occasionally attacking the source\cite{regnier1968insect,Dorigo06antcolony}. Wilson et al. defined a specific behavior of \textit{Solenopsis invicta} Buren, namely alarm behavior, under panic where individuals move in a rapid and unstable pattern\cite{Wilson1962}. Based on these biological facts, we proposed alarm pheromone model\cite{Li2014} with simple rules: (1) pheromone will be left where the individual resides, (2) pheromone will evaporate with time, (3) the moving direction of the agent at each time step is determined by combination of current direction and the direction pointing to local pheromone with maximum concentration, (4) agent cannot distinguish the direction of pheromone source if more than one source with concentration higher than the detection threshold. We firstly use this model to simulate the effect of initially asymmetric distribution, characterized by Initial Difference, on symmetry breaking. The moving area of agents in model is divided into left and right part equally. Initial Difference can be defined as the ratio difference in amount of ants between the left and right part(Fig. \ref{fig:Fig_1}):

\begin{equation}
    Initial\; Difference = \frac{\left|\;total\; number\; of\; ants\; of\; left\; part\; -\; total\; number\; of\; ants\; of\; left\; part\;\right|}{total\; number\; of\; ants}
\end{equation}

and is set to 0, 0.2, 0.5, 0.7, 1.0 corresponding to \{10,10\},\{12,8\},\{15,5\},\{17,3\},\{0,20\}, namely the number of ants in left and right part.Following the methods of our previous work, we use Collective Asymmetry(CA) to characterize symmetry breaking induced by panic\cite{Li2014}(see Methods section):

\begin{equation}
    Collective\; Asymmetry = difference\; -\; random\; difference
\end{equation}

\begin{figure}[ht]
\centering
\includegraphics[width=\linewidth]{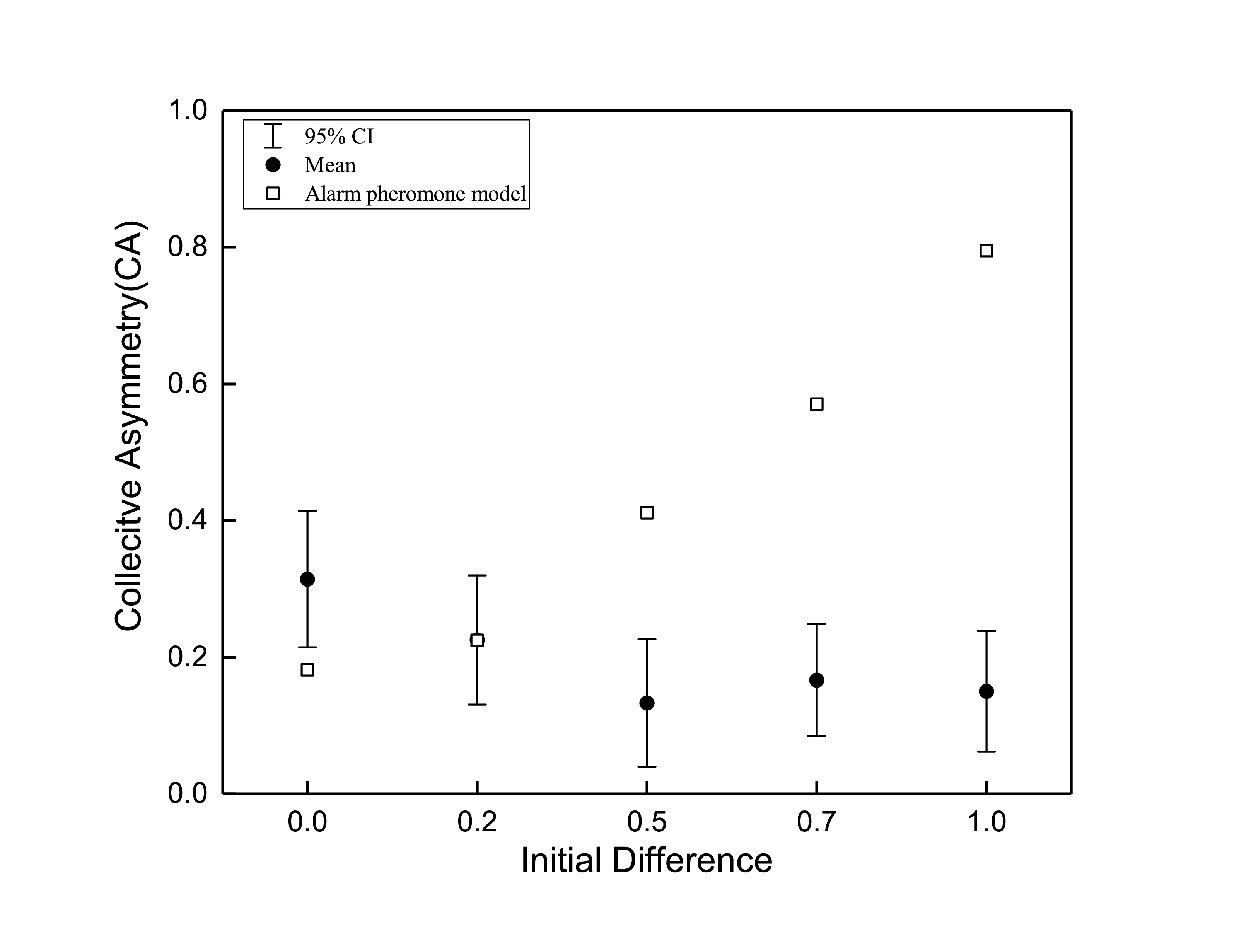}
\caption{\textbf{The comparison of the alarm pheromone model and the current experiment results.} The hollow square is the result averaged over 10,000 runs of the alarm pheromone model\cite{Li2014}. The average of CA of the current experiment over 30 repetitions is shown by the solid circle and the whiskers extend to 95\% confidential interval (CI). Obviously, the alarm pheromone model shows an increasing trend with the increase of Initial Difference, thus far, cannot reproduce the current experiment results.}
\label{fig:Fig_2}
\end{figure}

The simulation result of alarm pheromone model shows an obviously positive correlation between CA and Initial Difference (Fig. \ref{fig:Fig_2}, hollow square), which is in line with intuition. To validate the simulation result, it is necessary to perform the panic escaping experiment with ant colonies. We designed a circular cell with two symmetrically located exits, which were initially blocked so that ants could not escape (Fig. \ref{fig:Fig_1}). At the center of the cell, a dose of 50$\mu$l citronella (a kind of repellent) was injected to induce panic. Same as the model simulation, the configurations of the number of ants are {number of left side, number of right side} = \{10, 10\}, \{12, 8\}, \{15, 5\}, \{17, 3\} and \{0, 20\}, corresponding to Initial Difference = 0, 0.2, 0.5, 0.7, 1.0. 30 repetitions were performed for each Initial Difference under the same controlled condition. Surprisingly, the experiment results show no positive correlation between Initial Difference and CA (Fig. \ref{fig:Fig_2}, solid circle with whiskers), which is counter-intuitive and quite different from that of the simulation. The alarm pheromone model cannot reproduce the experiment results. Hence, some adjustments are needed on its mechanisms based on deeper investigation of current experiment.

\subsection*{Flee-pheromone model}

\begin{figure}[ht]
\centering
\includegraphics[width=\linewidth]{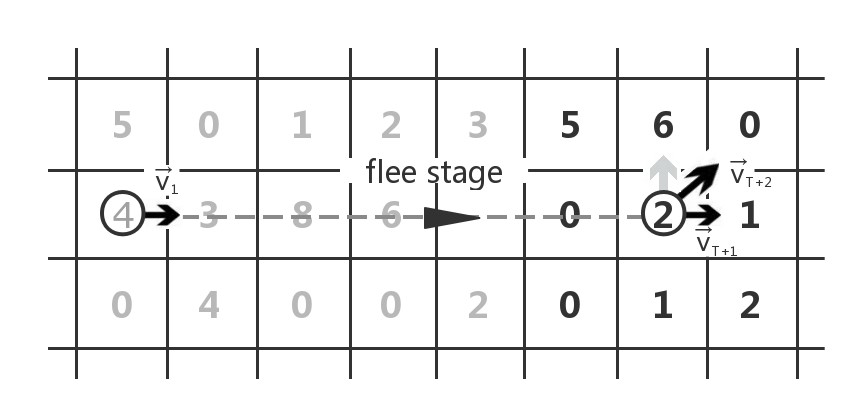}
\caption{\textbf{Schematic diagram demonstrating the flee-pheromone model.} This panel shows the updating rules of the flee-pheromone model from time step 1 to $T + 1$, namely from the flee stage to the onset of the pheromone stage, where $T$ is the duration of flee stage. The hollow circle denotes one agent, and the black arrow denotes its velocity vector, shown as $\vec V_{i}$ at time step $i$. The numbers on the lattice denote the amount of pheromone. The dashed grey line shows the trajectory of an agent during the flee stage, along the initial direction randomly selected denoted by the black triangle. Lattices with light grey numbers mean that the ant’s behavior is not regulated by pheromone on them during the flee stage. At time step $T+1$, pheromone stage starts and the moving direction is denoted by $\vec V_{T+1}$. The gray arrow pointing to the lattice labeled with “6” denotes the moving propensity to the cell with largest concentration of pheromone within detection range. The moving direction of time step $\vec V_{T+2}$ is thus the combination of propensity direction and moving direction at $T+1$ according to formula (3), pointing to the lattice labeled by “0”. It should be noticed that in simulation, the amount of pheromone an agent puts on the lattice and the amount of pheromone evaporates each time step are both less than 1. For easy understanding, we show integers here.}
\label{fig:Fig_3}
\end{figure}

The failure of the alarm pheromone model in explaining the current experiment reminded us some details may be neglected. We thus further analyzed ant’s panic escape and found a special behavior. Different from the moving patterns regulated by alarm pheromone we assumed in the previous model, ants would move "relatively straight" with orientation randomly chosen. Noting that the actual trajectories of ants are not strictly straight, "relatively straight" used here means that smooth arcs and broken lines with small tortuosity would be seen as "moving straight". This special behavior can be observed and last for a short period of time just at the beginning. It seems that the alarm pheromone is not a major factor that determines ant’s behavior during this special stage. We define such a special stage as “flee stage”. After flee stage, ant’s behavior is assumed to be determined by pheromone, the same with alarm pheromone model. Hence the later stage is defined as “pheromone stage”. Therefore, flee-pheromone model can be constructed with the combination of these two stages sequentially in time. Following the simple form of cellular automatic model, rules of flee-pheromone model are described as follows with a schematic diagram presented in Fig. \ref{fig:Fig_3}:

1)Agents can move in a two-dimensional square chamber with $L\times L$ lattices in total. Initially, number of N agents are distributed in the chamber according to Initial Difference, each with a moving direction randomly chosen. The detection range of each ant is its nearest 8 lattices.

2)Flee stage. Define $T$ as the duration of the flee stage. Each individual moves straight forward along the initial direction from time step 1 to $T$ and reflects if meets the boundary. It should be noticed that behaviors of ants are quite variable and are simplified here in model. 

3)Pheromone stage. After flee stage, agents would update the position with the combination of the direction to lattice with the largest amount of pheromone within detection range and current moving direction, according to the equation\cite{Li2014}:

\begin{equation}
\left\{
\begin{aligned}
x(t)\; & = & \; x(t-1)\; +\; \{\; v_x\; +\; p_x^m\;\} \\
y(t)\;& = & \; y(t-1)\; +\; \{\; v_y\; +\; p_y^m\;\} \\
\end{aligned}
\right.
\end{equation}

Here, $x(t)$ and $y(t)$ denote the $x$ and $y$ components of the position vector at time step $t$, respectively. $t-1$ denotes the one previous time step from time step $t$. $v_x$ and $v_y$ denote the $x$ and $y$ components of the velocity vector, respectively, which can only take the values of 1, 0 and -1. In addition, 0 cannot be assigned for both $v_x$ and $v_y$ simultaneously. $p_m$ denotes the vector pointing to the lattice with maximum amount of pheromone, and $p_x^m$ and $p_y^m$ are the x and y components. The bracket $\{\alpha\}$ is used to regulate the moving speed, ensuring each agent moves exactly one lattice in every time step, defined as

\begin{equation}
 \{\alpha\}=\left\{
\begin{aligned}
1, \;if\;\alpha>0 \\
0, \;if\;\alpha=0 \\
-1, \;if\;\alpha<0
\end{aligned}
\right.   
\end{equation}
 
 Considering the limited sensing ability of agents, if the detected amount of pheromone is larger than the given threshold $p_t$, it is treated as $p_t$. Meanwhile, if there are more than one lattice with the pheromone amount equal or larger than $p_t$, agents would select a lattice randomly. 
 
Specifically, each time step a constant amount of pheromone $p_0$ would be deposited on the lattice an agent resides and can be accumulated in both stages. The amount of pheromone on each lattice would evaporate with a constant rate $\Delta p$ until it reaches zero. Agents would escape from the chamber once an exit is detected and reflect when meeting the boundary. The values of the main parameters are shown in the Methods section.

\begin{figure}[ht]
\centering
\includegraphics[width=\linewidth]{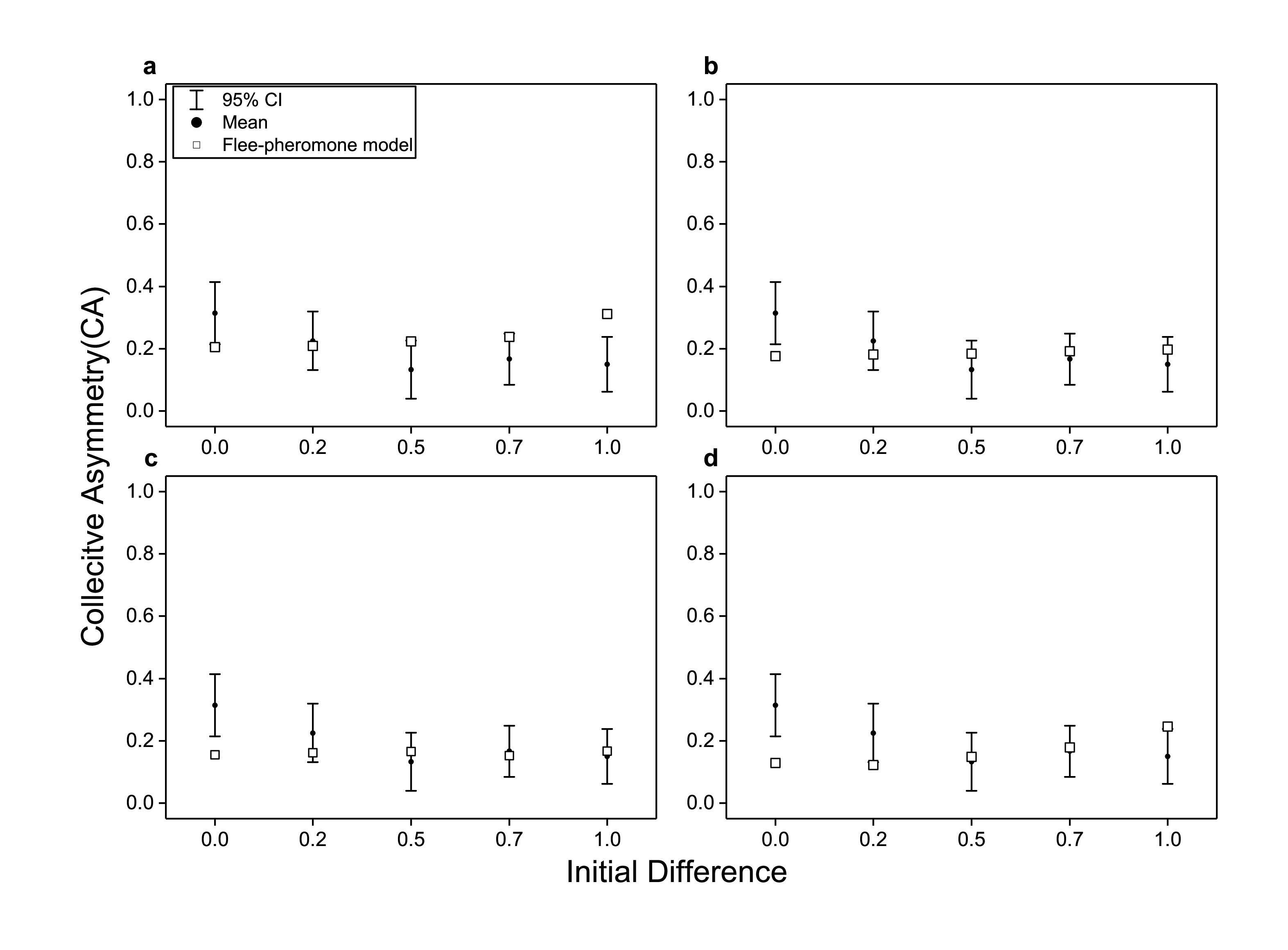}
\caption{\textbf{The comparison of the flee-pheromone model and the current experiment results.} The hollow square denotes result of flee-pheromone model averaged over 10,000 runs, the average of CA of the current experiment over 30 repetitions is shown by the solid circle and whiskers extend to 95\% confidential interval (CI). Flee stage duration is set to be 8 to 16 time steps respectively, corresponding to 3.32 to 6.63 seconds in real world time. Only duration 8, 10, 12, 16 respectively for (a), (b), (c) and (d) are shown as illustration (the duration from 6 to 17 time steps can be found as Supplementary Fig. S1). The model quantitatively reproduces the experiment results.}
\label{fig:Fig_4}
\end{figure}

We find that when the flee stage duration $T$ falls in the range 8 to 16 time steps, corresponding to about 3 to 7 seconds of real world time, flee-stage pheromone model fits well with that of the experiment both qualitatively and quantitatively (Fig. \ref{fig:Fig_4}). As the ants’ average speed is measured to be 0.94 cm/s, the range of moving distance of ants is about 3 to 6 cm, close to 4 cm, radius of the cell used in the experiment, meaning that ants can travel from one half part of the cell to another. This demonstrates that the flee stage alleviates the unevenness of the initial distribution and is likely the key factor leading to the unexpected experiment results.

\subsection*{Calibration of the flee stage duration}

A theoretical model that attempts to reveal the underlying mechanism regulating the ants’ behavior should demonstrate its power by not only fitting the current experiment but also explaining the previously related experiments. The current model in this study is in such a case. Our previous work demonstrated that ant colony displays an increase then decrease of symmetry breaking with ant density increases\cite{Li2014}. This experiment was performed without considering any biased initial distributions, so whether the flee-pheromone model can explain the experiment with the same duration of flee stage needs to be validated; in other word, this previous experiment results can be used to calibrate the flee pheromone model, specifically for flee stage time to get a general model that explains both experiments. 

\begin{figure}[ht]
\centering
\includegraphics[width=\linewidth]{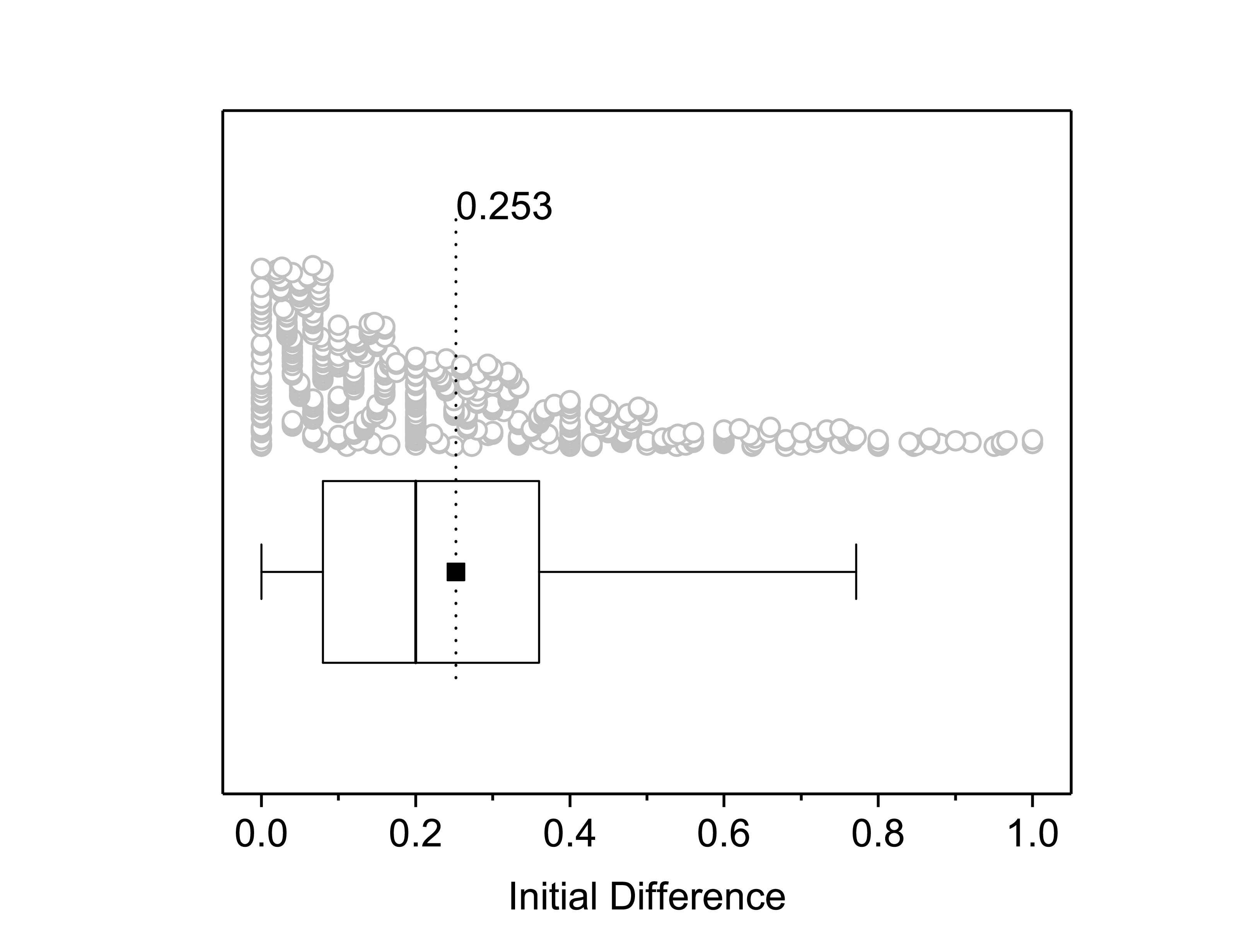}
\caption{\textbf{The distribution of Initial Difference of experiments\cite{Li2014} on density and symmetry breaking.} The light grey circle shows the distribution of Initial Difference of the previous experiment at the onset of escaping. The box chart on the bottom shows statistic features. The average of Initial Distribution is 0.253 and is shown by the solid square. The median is shown by the solid line. The interquartile range is enclosed by the box. The whiskers extend to the most extreme data point within 1.5$\ times$ the interquartile range outside the box. The frequency distribution on the top shows that the major part of the previous experiment\cite{Li2014} takes a relatively symmetric distribution with low Initial Difference.}
\label{fig:Fig_5}
\end{figure}

\begin{figure}[ht]
\centering
\includegraphics[width=\linewidth]{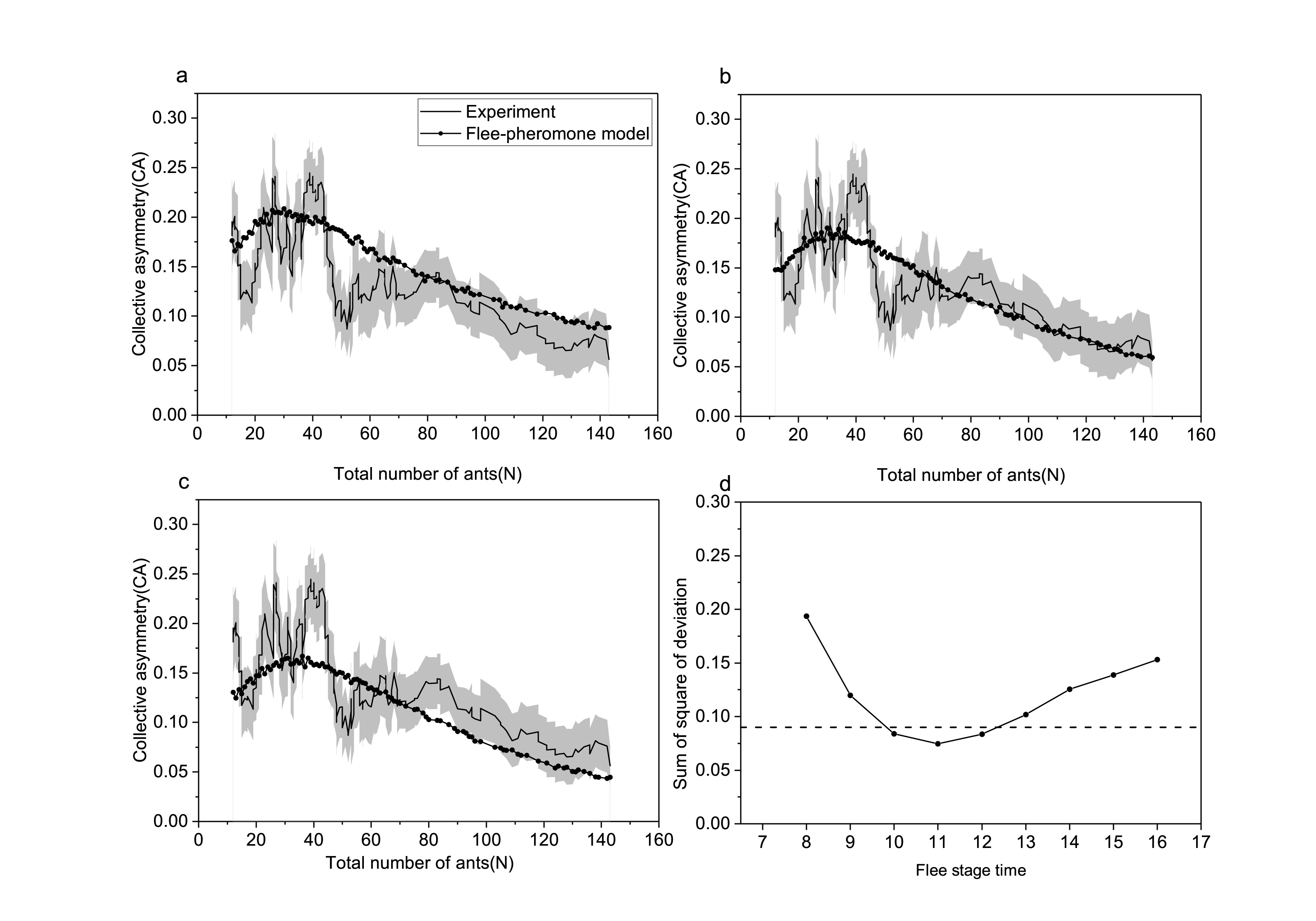}
\caption{\textbf{The comparison of flee-pheromone model and experiment results demonstrated by our previous work in 2014\cite{Li2014}.} The moving average of CA over 31 points is shown by lines and the standard error of the average is shown by the grey area. The model result is shown by solid circle over the average of 5,000 runs. The Initial Difference of the model is set to 0.253 as extracted from the experiment demonstrated in Fig. 5. The flee stage duration is set to be 8 to 16 time steps. Only the simulation results with flee stage duration 9, 11 and 13, corresponding to panel (a), (b) and (c), are shown in this figure as an illustration (the duration from 8 to 16 time steps can be found as Supplementary Fig. S2). The fitness for each flee stage duration with experiment is evaluated by the sum of square of deviation relative to average CA for each total number of ants N, as shown in panel (d). In panel (d), the horizontally solid line denotes 0.09, below which can be considered fitting well with the experiment, corresponding to flee stage duration 10, 11 and 12.}
\label{fig:Fig_6}
\end{figure}

By analyzing videos recording experiment procedures\cite{Li2014} through pattern recognition technology, we confirm that the Initial Difference is 0.253 on average (Fig. \ref{fig:Fig_5}), which means the experiment was carried out with relatively even distribution. Under this Initial Difference, flee-pheromone model was performed with flee stage duration $T$ ranging from 8 to 16 time steps. With the sum of square of deviation as an evaluation of fitness, flee-pheromone model nicely matches the density dependence of escaping symmetry breaking with the flee stage duration equals 10, 11 and 12(Fig. \ref{fig:Fig_6}, sum of square of deviation < 0.09) and at the same time, its reliability can be demonstrated in this case. we thus establish a uniform model that incorporates the original alarm pheromone model with a well generalization ability that explain both experiments.

\section*{Discussion}

Collective aggregation is a universal phenomenon in nature and can be classified according to the cause resulting in the assemble\cite{Jeanson2005}. On the one hand, aggregation can be the result of the reaction to heterogeneous factors of the environment, food resources as an example, which becomes a cue tempting animals to move towards it. Once the heterogeneous factors are removed, for example food is consumed, aggregation will not sustain and animals will disperse in the environment. On the other hand, aggregation can emerge from social interactions among individuals. As for animals with high sociality, collective aggregation based on self-organized process to become adaptive system is of vital importance to survive under various environmental pressures\cite{Bonabeau1997}. Both kinds of aggregation will lead to the uneven distribution of animals in space, essentially a form of symmetry breaking. 

It is phenomenally interesting and theoretically important to scrutinize whether the initially asymmetric distribution can influence symmetry breaking under panic. Considering the success in our previous work, alarm pheromone model was firstly used to make a prediction. The simulation showed a positive correlation between Initial Difference and Collective Asymmetry and was quite in line with intuitive, but to our surprise, obviously different from the current experiment. 

The comparison between model prediction and experiment not only further examined the correctness of alarm pheromone, but also provides an opportunity to discover the hidden mechanism. Based on analysis of ant panic escape, flee stage was proposed for the first time, bringing a totally new perspective both for empirical study and modelling method on ant panic escape. As for collective behavior of ants, differential equation method based on macro analysis of phenomenon and self-propelled particle (SPP) model focuses on micro interactions are two main methods widely used. It has been demonstrated that foraging behavior of ants can be separated into different stages according to behavior features, therefore, differential equations of every stage can be derived once the variance of the amount of ants with time is detected and give a well explanation on foraging behavior\cite{Bartholdi1993,Beckers1990,Beckers1993,Beekman2001,Bonabeau1996,Deneubourg1990,Sumpter2003}. Different from foraging, however, the research on panic escaping of ants concentrates little on the separation of different stages and lacks of empirical evidence to draw differential equations. On the other hand, SPP model has made progress on explaining the decision-making of ants at the bifurcation point of the road based on pheromone detection\cite{Perna2012,Beckers1992,Calenbuhr1992}. But still, as mentioned, ant under panic shows quite different behavior from that during space searching, so these existing models cannot be applied directly to explain the symmetry breaking induced by panic. 

Up to now, empirical research on ant panic escape mainly concentrates on macro features with statistical analysis\cite{Beekman2001,Boari2013,dias2012intersecting,Hu2003,Schwinghammer2006,Gautam2011,Wang2015} in experiments, however, an essential mechanism of whether there are different stages during escaping was not emphasized. This current work explicitly distinguished two stages based on flee stage assumption that preliminarily verified by current and previous experiment, potentially providing an idea for stage division in ant panic escape. With flee-pheromone model proposed in this paper as a reference, more studies on panic escape can be performed to further demonstrate the assumption of two-stage division on panic escape.

As an intuitive explanation for the counter-intuitive phenomenon of current experiment, there is little alarm pheromone deposed in space to guide individuals’ moving at the beginning of panic escape, ants thus may prefer to move straight forward in a random direction. Such a moving pattern breaks the initially asymmetry distribution and makes the distribution of ants random and even. Actually, it is discovered that fish schooling would escape from all directions once predators are detected\cite{Nakamura1972}, just like the flee stage for \textit{Solenopsis invicta} Buren. Hence, "flee stage" may be a universal pattern remains to be investigated across various animal species. This shared feature, even if may emerge from different interaction rules between individuals, can be a inspiration for research on collective behavior. One natural idea is that individual behavior may be regulated by different rules in different stages, and more importantly, the transition between different stage can be a result of individuals' interaction with local environment and the conspecific. Rather than adopting one static rule, behavioral division introduces some dynamic factors in modelling collective behavior and can be more close to biological facts.

While the flee-pheromone model successfully reproduced the current experiment, we still know little about the way that ant releases and detects alarm pheromone during flee stage, more empirical studies on panic experiment are necessary to illustrate the details. Sumpter et al. proposed a method based on “new imaging and analysis technology” to estimate pheromone trail during space searching of ants\cite{Perna2012}. Inspired by this idea, we propose that reconstructing the pheromone releasing and detecting process by image analysis of the ant behavior may potentially provide a way to quantitatively investigate the transition point between the flee stage and the pheromone stage. Another work recently published also provides an idea for stage division. In this work\cite{Tejera2016}, Tejera et al. demonstrated that for ants in foraging danger information cannot be transmitted along long distances and proposed different kinds of stimulus may cause different panic response; it indicates that the transition among different stages can possibly be determined by types of panic source. Related investigation thus can be performed accordingly.

\section*{Methods}

\subsection*{Experimental subjects}
Hundreds of red imported ant (\textit{Solenopsis invicta} Buren) workers were collected from a single nest in Jiaying University located in Meizhou city , Guangdong province, China in May, 2017 and were fed in cell. The cell was about 30 mm in length and 20 mm in width coated with Fluon to prevent ants from escaping. The ants were fed several weeks with Tenebrio molitor before the experiments. The temperature was held at 23 $^{\circ}$C.

\subsection*{Experimental protocol}

Trials were carried out in a circular cell made by glass with 8 cm in diameter, 0.5 cm in height and two exits, 1 cm in width, symmetrically located to the left and the right which were initially blocked. The cell was rested on several layers, including a piece of thin plastic yellow paper, a piece of clipped A4 paper and a piece of filter paper from top to bottom. Two desk lamps were placed beside the box symmetrically to maintain a closely uniform light intensity and thus prevented a possible moving direction preference induced by uneven light intensity.

For each trial, 20$\pm$2 ants with similar size were randomly chosen and introduced into two test tubes. The amount of ants for each test tube was determined by the value of Initial Difference. Test tubes were coated with Fluon so that the ants were not able to climb up and could be transferred easily.

Before introducing the ants, a dose of 50 $\mu L$ repelling liquid (citronella, Yanggongfang, China) was injected on the plastic yellow paper at the center. A few seconds later, the ants in the test tubes were all introduced into the left and right side of the cell, symmetrically to the center, and then the cell was covered by a 0.3 cm thick plastic cover immediately. The two exits were then opened simultaneously so that the ants were able to escape. The time between picking up the ants to opening the exits was typically approximately 30 seconds. The whole process was recorded by a video camera situated above the cell. 

There were 30 repetitions for each value of Initial Difference and 150 trials were performed in total. In each trial, a new group of 20$\pm$2 ants and new layers under the cell (including the filtering paper, plastic paper and A4 paper) were used to avoid liquid and possible pheromone residue in the equipment.

\subsection*{Definition of symmetry breaking}
Considering an $N$ particle random SPP model, for an unbiased movement, the possibility of escaping left and right for each particle should be equal, that is 1/2. After some simple statistical derivations, we obtain the random difference produced from an N particle random SPP model as

\begin{equation}
   random\ difference = \sum_{i=0}^{N}p(i)\cdot difference(i) = \sum_{i=0}^{N} \frac{C_{N}^{i}}{2^{N}}\cdot \frac{|i-(N-i)|}{N} 
\end{equation}

where $i$ denotes the total number of ants escaping left. $N$ denotes the total number of escaping ants. $p(i)$ and $difference(i)$ denote the possibility and difference that the total of ants escaping left is $i$, respectively.

\subsection*{Parameters of flee-pheromone model}
In the simulation, the following set of parameters is used (same with alarm pheromone model\cite{Li2014}): $L = 20$, $p_t = 1$, $p_0 = 1/5$ and $t_fo = 30$ seconds. We set $L$ always to be 20, so that it corresponds to the ratio between the size of the cell (8.0 cm) and the average body length (0.39 cm) of the ants we used, which was determined by video measurement from our experiment. That is to say, one ant occupies one lattice exactly. Considering that the absolute values of $p_t$ and $p_0$ are meaningless, and only the relative values make sense, we set the threshold $p_t$ always to be 1 as a reference value. To compare the fade-out time of the pheromone in the simulation with that in the real world, we introduce $t_{fo}$ to denote the fade-out time from $p_0$ to 0. As the ants average speed is measured to be 0.94 cm/s, the relationship among $t_{fo}$, $\Delta p$ and $p_0$ can be deduced as follows,

\begin{equation}
    t_{fo}\approx \frac{0.42p_0}{\Delta p}
\end{equation}

\subsection*{Pattern recognition and data analysis}

To obtain the Initial Difference of our previous experiment\cite{Li2014}, we used Matlab(2015b) to deal with videos recording experiment procedures. According to the definition of Initial Difference, only the number of ants in left and right side of the cell at the onset of each experiments are needed without tracking of each individuals along time, the pattern recognition used here is relatively simple. The main process is to obtain the first frame in each video and make it to be gray scale. Then, adjusting the threshold of gray scale to recognize the ants which are black compared to the background. The pixel number of ants in left and right of the cell is calculated to estimate the number of ants respectively. Totally 485 videos are analyzed. The statistic analysis of the both experiments was performed by OriginPro 2017.

\section*{Data Availability}

All data analyzed during this study are included in Supplementary Information files.

\section*{Acknowledgements}

The authors acknowledge funding supported by Scientific Research Department, Jiaying University (131/316E43), Students' Platform for Innovation and Entrepreneurship Training Program (201810582195, 201810582065) and National Natural Science Foundation of China (61374165). Weijia Wang thanks Meijun Wu (School of Systems Science, Beijing Normal University) for comment on the manuscript.

\section*{Author contributions statement}

G.L., W.J.W, H.B.W., J.P.W., Z.R.D., B.R. and Z.G.H. designed the study, J.H.L. and Z.Y.H. carried out the experiment, J.Q.L. carried out pattern recognition, W.J.W. analyzed the data, G.L. and W.J.W. constructed the model, W.J.W. performed simulation, G.L., W.J.W. and Z.G.H. wrote the paper. 

\section*{Additional information}

\textbf{Competing interests:} The authors declare no competing interests.

\begin{thebibliography}{10}
\urlstyle{rm}
\expandafter\ifx\csname url\endcsname\relax
  \def\url#1{\texttt{#1}}\fi
\expandafter\ifx\csname urlprefix\endcsname\relax\def\urlprefix{URL }\fi
\expandafter\ifx\csname doiprefix\endcsname\relax\def\doiprefix{DOI: }\fi
\providecommand{\bibinfo}[2]{#2}
\providecommand{\eprint}[2][]{\url{#2}}

\bibitem{Altshuler2005}
\bibinfo{author}{Altshuler, E.} \emph{et~al.}
\newblock \bibinfo{journal}{\bibinfo{title}{Symmetry breaking in escaping
  ants}}.
\newblock {\emph{\JournalTitle{The American Naturalist}}}
  \textbf{\bibinfo{volume}{166}}, \bibinfo{pages}{643--649},
  \doiprefix\url{10.1086/498139} (\bibinfo{year}{2005}).

\bibitem{Boari2013}
\bibinfo{author}{Boari, S.}, \bibinfo{author}{Josens, R.} \&
  \bibinfo{author}{Parisi, D.~R.}
\newblock \bibinfo{journal}{\bibinfo{title}{Efficient egress of escaping ants
  stressed with temperature}}.
\newblock {\emph{\JournalTitle{{PLoS} {ONE}}}} \textbf{\bibinfo{volume}{8}},
  \bibinfo{pages}{e81082}, \doiprefix\url{10.1371/journal.pone.0081082}
  (\bibinfo{year}{2013}).

\bibitem{Parisi2015}
\bibinfo{author}{Parisi, D.}, \bibinfo{author}{Soria, S.} \&
  \bibinfo{author}{Josens, R.}
\newblock \bibinfo{journal}{\bibinfo{title}{Faster-is-slower effect in escaping
  ants revisited: Ants do not behave like humans}}.
\newblock {\emph{\JournalTitle{Safety Science}}} \textbf{\bibinfo{volume}{72}},
  \bibinfo{pages}{274--282}, \doiprefix\url{10.1016/j.ssci.2014.09.014}
  (\bibinfo{year}{2015}).

\bibitem{Soria2012}
\bibinfo{author}{Soria, S.}, \bibinfo{author}{Josens, R.} \&
  \bibinfo{author}{Parisi, D.}
\newblock \bibinfo{journal}{\bibinfo{title}{Experimental evidence of the
  {\textquotedblleft}faster is slower{\textquotedblright} effect in the
  evacuation of ants}}.
\newblock {\emph{\JournalTitle{Safety Science}}} \textbf{\bibinfo{volume}{50}},
  \bibinfo{pages}{1584--1588}, \doiprefix\url{10.1016/j.ssci.2012.03.010}
  (\bibinfo{year}{2012}).

\bibitem{Chung2017}
\bibinfo{author}{Chung, Y.-K.} \& \bibinfo{author}{Lin, C.-C.}
\newblock \bibinfo{journal}{\bibinfo{title}{Heat-induced symmetry breaking in
  ant (hymenoptera: Formicidae) escape behavior}}.
\newblock {\emph{\JournalTitle{{PLOS} {ONE}}}} \textbf{\bibinfo{volume}{12}},
  \bibinfo{pages}{e0173642}, \doiprefix\url{10.1371/journal.pone.0173642}
  (\bibinfo{year}{2017}).

\bibitem{Saloma11947}
\bibinfo{author}{Saloma, C.}, \bibinfo{author}{Perez, G.~J.},
  \bibinfo{author}{Tapang, G.}, \bibinfo{author}{Lim, M.} \&
  \bibinfo{author}{Palmes-Saloma, C.}
\newblock \bibinfo{journal}{\bibinfo{title}{Self-organized queuing and
  scale-free behavior in real escape panic}}.
\newblock {\emph{\JournalTitle{Proceedings of the National Academy of
  Sciences}}} \textbf{\bibinfo{volume}{100}}, \bibinfo{pages}{11947--11952},
  \doiprefix\url{10.1073/pnas.2031912100} (\bibinfo{year}{2003}).

\bibitem{Saloma2015}
\bibinfo{author}{Saloma, C.}, \bibinfo{author}{Perez, G.~J.},
  \bibinfo{author}{Gavile, C.~A.}, \bibinfo{author}{Ick-Joson, J.~J.} \&
  \bibinfo{author}{Palmes-Saloma, C.}
\newblock \bibinfo{journal}{\bibinfo{title}{Prior individual training and
  self-organized queuing during group emergency escape of mice from water
  pool}}.
\newblock {\emph{\JournalTitle{{PLOS} {ONE}}}} \textbf{\bibinfo{volume}{10}},
  \bibinfo{pages}{e0118508}, \doiprefix\url{10.1371/journal.pone.0118508}
  (\bibinfo{year}{2015}).

\bibitem{Helbing2000}
\bibinfo{author}{Helbing, D.}, \bibinfo{author}{Farkas, I.} \&
  \bibinfo{author}{Vicsek, T.}
\newblock \bibinfo{journal}{\bibinfo{title}{Simulating dynamical features of
  escape panic}}.
\newblock {\emph{\JournalTitle{Nature}}} \textbf{\bibinfo{volume}{407}},
  \bibinfo{pages}{487--490}, \doiprefix\url{10.1038/35035023}
  (\bibinfo{year}{2000}).

\bibitem{Helbing2002}
\bibinfo{author}{Helbing, D.}, \bibinfo{author}{Farkas, I.},
  \bibinfo{author}{Molnar, P.} \& \bibinfo{author}{Vicsek, T.}
\newblock \emph{\bibinfo{title}{Simulation of pedestrian crowds in normal and
  evacuation situations}}, vol.~\bibinfo{volume}{21}, \bibinfo{pages}{21--58}
  (\bibinfo{publisher}{Springer}, \bibinfo{year}{2002}).

\bibitem{Helbing2005}
\bibinfo{author}{Helbing, D.}, \bibinfo{author}{Buzna, L.},
  \bibinfo{author}{Johansson, A.} \& \bibinfo{author}{Werner, T.}
\newblock \bibinfo{journal}{\bibinfo{title}{Self-organized pedestrian crowd
  dynamics: Experiments, simulations, and design solutions}}.
\newblock {\emph{\JournalTitle{Transportation Science}}}
  \textbf{\bibinfo{volume}{39}}, \bibinfo{pages}{1--24},
  \doiprefix\url{10.1287/trsc.1040.0108} (\bibinfo{year}{2005}).

\bibitem{Li2014}
\bibinfo{author}{Li, G.} \emph{et~al.}
\newblock \bibinfo{journal}{\bibinfo{title}{Symmetry breaking on density in
  escaping ants: Experiment and alarm pheromone model}}.
\newblock {\emph{\JournalTitle{{PLoS} {ONE}}}} \textbf{\bibinfo{volume}{9}},
  \bibinfo{pages}{e114517}, \doiprefix\url{10.1371/journal.pone.0114517}
  (\bibinfo{year}{2014}).

\bibitem{Ji2018}
\bibinfo{author}{Ji, Q.}, \bibinfo{author}{Xin, C.}, \bibinfo{author}{Tang, S.}
  \& \bibinfo{author}{Huang, J.}
\newblock \bibinfo{journal}{\bibinfo{title}{Symmetry associated with symmetry
  break: Revisiting ants and humans escaping from multiple-exit rooms}}.
\newblock {\emph{\JournalTitle{Physica A: Statistical Mechanics and its
  Applications}}} \textbf{\bibinfo{volume}{492}}, \bibinfo{pages}{941--947},
  \doiprefix\url{10.1016/j.physa.2017.11.024} (\bibinfo{year}{2018}).

\bibitem{Wang2016}
\bibinfo{author}{Wang, S.}, \bibinfo{author}{Cao, S.}, \bibinfo{author}{Wang,
  Q.}, \bibinfo{author}{Lian, L.} \& \bibinfo{author}{Song, W.}
\newblock \bibinfo{journal}{\bibinfo{title}{Effect of exit locations on ants
  escaping a two-exit room stressed with repellent}}.
\newblock {\emph{\JournalTitle{Physica A: Statistical Mechanics and its
  Applications}}} \textbf{\bibinfo{volume}{457}}, \bibinfo{pages}{239--254},
  \doiprefix\url{10.1016/j.physa.2016.03.083} (\bibinfo{year}{2016}).

\bibitem{Portha2002}
\bibinfo{author}{Portha, S.}
\newblock \bibinfo{journal}{\bibinfo{title}{Self-organized asymmetries in ant
  foraging: a functional response to food type and colony needs}}.
\newblock {\emph{\JournalTitle{Behavioral Ecology}}}
  \textbf{\bibinfo{volume}{13}}, \bibinfo{pages}{776--781},
  \doiprefix\url{10.1093/beheco/13.6.776} (\bibinfo{year}{2002}).

\bibitem{Price2016}
\bibinfo{author}{Price, R.~I.}, \bibinfo{author}{Grüter, C.},
  \bibinfo{author}{Hughes, W. O.~H.} \& \bibinfo{author}{Evison, S. E.~F.}
\newblock \bibinfo{journal}{\bibinfo{title}{Symmetry breaking in
  mass-recruiting ants: extent of foraging biases depends on resource
  quality}}.
\newblock {\emph{\JournalTitle{Behavioral Ecology and Sociobiology}}}
  \textbf{\bibinfo{volume}{70}}, \bibinfo{pages}{1813--1820},
  \doiprefix\url{10.1007/s00265-016-2187-y} (\bibinfo{year}{2016}).

\bibitem{Wilson1971}
\bibinfo{author}{Wilson, E.~O.} \& \bibinfo{author}{Regnier, F.~E.}
\newblock \bibinfo{journal}{\bibinfo{title}{The evolution of the alarm-defense
  system in the formicine ants}}.
\newblock {\emph{\JournalTitle{The American Naturalist}}}
  \textbf{\bibinfo{volume}{105}}, \bibinfo{pages}{279--289},
  \doiprefix\url{10.1086/282724} (\bibinfo{year}{1971}).

\bibitem{Attanasi2014}
\bibinfo{author}{Attanasi, A.} \emph{et~al.}
\newblock \bibinfo{journal}{\bibinfo{title}{Collective behaviour without
  collective order in wild swarms of midges}}.
\newblock {\emph{\JournalTitle{{PLoS} Computational Biology}}}
  \textbf{\bibinfo{volume}{10}}, \bibinfo{pages}{e1003697},
  \doiprefix\url{10.1371/journal.pcbi.1003697} (\bibinfo{year}{2014}).

\bibitem{Attanasi2014a}
\bibinfo{author}{Attanasi, A.} \emph{et~al.}
\newblock \bibinfo{journal}{\bibinfo{title}{Finite-size scaling as a way to
  probe near-criticality in natural swarms}}.
\newblock {\emph{\JournalTitle{Physical Review Letters}}}
  \textbf{\bibinfo{volume}{113}},
  \doiprefix\url{10.1103/physrevlett.113.238102} (\bibinfo{year}{2014}).

\bibitem{Allee1927}
\bibinfo{author}{Allee, W.~C.}
\newblock \bibinfo{journal}{\bibinfo{title}{Animal aggregations}}.
\newblock {\emph{\JournalTitle{The Quarterly Review of Biology}}}
  \textbf{\bibinfo{volume}{2}}, \bibinfo{pages}{367--398},
  \doiprefix\url{10.1086/394281} (\bibinfo{year}{1927}).

\bibitem{Parrish1999}
\bibinfo{author}{Parrish, J.~K.}
\newblock \bibinfo{journal}{\bibinfo{title}{Complexity, pattern, and
  evolutionary trade-offs in animal aggregation}}.
\newblock {\emph{\JournalTitle{Science}}} \textbf{\bibinfo{volume}{284}},
  \bibinfo{pages}{99--101}, \doiprefix\url{10.1126/science.284.5411.99}
  (\bibinfo{year}{1999}).

\bibitem{JohnT.Emlen1952}
\bibinfo{author}{Emlen, J.~T.}
\newblock \bibinfo{journal}{\bibinfo{title}{Flocking behavior in birds}}.
\newblock {\emph{\JournalTitle{The Auk}}} \textbf{\bibinfo{volume}{69}},
  \bibinfo{pages}{160--170}, \doiprefix\url{10.2307/4081266}
  (\bibinfo{year}{1952}).

\bibitem{Ballerini2008}
\bibinfo{author}{Ballerini, M.} \emph{et~al.}
\newblock \bibinfo{journal}{\bibinfo{title}{Empirical investigation of starling
  flocks: a benchmark study in collective animal behaviour}}.
\newblock {\emph{\JournalTitle{Animal Behaviour}}}
  \textbf{\bibinfo{volume}{76}}, \bibinfo{pages}{201--215},
  \doiprefix\url{10.1016/j.anbehav.2008.02.004} (\bibinfo{year}{2008}).

\bibitem{Cole1996}
\bibinfo{author}{Cole, B.~J.} \& \bibinfo{author}{Cheshire, D.}
\newblock \bibinfo{journal}{\bibinfo{title}{Mobile cellular automata models of
  ant behavior: Movement activity of leptothorax allardycei}}.
\newblock {\emph{\JournalTitle{The American Naturalist}}}
  \textbf{\bibinfo{volume}{148}}, \bibinfo{pages}{1--15},
  \doiprefix\url{10.1086/285908} (\bibinfo{year}{1996}).

\bibitem{Theraulaz2002}
\bibinfo{author}{Theraulaz, G.} \emph{et~al.}
\newblock \bibinfo{journal}{\bibinfo{title}{Spatial patterns in ant colonies}}.
\newblock {\emph{\JournalTitle{Proceedings of the National Academy of
  Sciences}}} \textbf{\bibinfo{volume}{99}}, \bibinfo{pages}{9645--9649},
  \doiprefix\url{10.1073/pnas.152302199} (\bibinfo{year}{2002}).

\bibitem{Mlot2011}
\bibinfo{author}{Mlot, N.~J.}, \bibinfo{author}{Tovey, C.~A.} \&
  \bibinfo{author}{Hu, D.~L.}
\newblock \bibinfo{journal}{\bibinfo{title}{Fire ants self-assemble into
  waterproof rafts to survive floods}}.
\newblock {\emph{\JournalTitle{Proceedings of the National Academy of
  Sciences}}} \textbf{\bibinfo{volume}{108}}, \bibinfo{pages}{7669--7673},
  \doiprefix\url{10.1073/pnas.1016658108} (\bibinfo{year}{2011}).

\bibitem{Vernerey2018}
\bibinfo{author}{Vernerey, F.~J.}, \bibinfo{author}{Shen, T.},
  \bibinfo{author}{Sridhar, S.~L.} \& \bibinfo{author}{Wagner, R.~J.}
\newblock \bibinfo{journal}{\bibinfo{title}{How do fire ants control the
  rheology of their aggregations? a statistical mechanics approach}}.
\newblock {\emph{\JournalTitle{Journal of The Royal Society Interface}}}
  \textbf{\bibinfo{volume}{15}}, \bibinfo{pages}{20180642},
  \doiprefix\url{10.1098/rsif.2018.0642} (\bibinfo{year}{2018}).

\bibitem{Nordlund1976}
\bibinfo{author}{Nordlund, D.~A.} \& \bibinfo{author}{Lewis, W.~J.}
\newblock \bibinfo{journal}{\bibinfo{title}{Terminology of chemical releasing
  stimuli in intraspecific and interspecific interactions}}.
\newblock {\emph{\JournalTitle{Journal of Chemical Ecology}}}
  \textbf{\bibinfo{volume}{2}}, \bibinfo{pages}{211--220},
  \doiprefix\url{10.1007/bf00987744} (\bibinfo{year}{1976}).

\bibitem{Wilson1962}
\bibinfo{author}{Wilson, E.~O.}
\newblock \bibinfo{journal}{\bibinfo{title}{Chemical communication among
  workers of the fire ant solenopsis saevissima (fr. smith) 1. the organization
  of mass-foraging}}.
\newblock {\emph{\JournalTitle{Animal Behaviour}}}
  \textbf{\bibinfo{volume}{10}}, \bibinfo{pages}{134--147},
  \doiprefix\url{10.1016/0003-3472(62)90141-0} (\bibinfo{year}{1962}).

\bibitem{Cronin2012}
\bibinfo{author}{Cronin, A.~L.}
\newblock \bibinfo{journal}{\bibinfo{title}{Consensus decision making in the
  ant myrmecina nipponica: house-hunters combine pheromone trails with quorum
  responses}}.
\newblock {\emph{\JournalTitle{Animal Behaviour}}}
  \textbf{\bibinfo{volume}{84}}, \bibinfo{pages}{1243--1251},
  \doiprefix\url{10.1016/j.anbehav.2012.08.036} (\bibinfo{year}{2012}).

\bibitem{HOLLDOBLER1976}
\bibinfo{author}{HOLLDOBLER, B.}
\newblock \bibinfo{journal}{\bibinfo{title}{Tournaments and slavery in a desert
  ant}}.
\newblock {\emph{\JournalTitle{Science}}} \textbf{\bibinfo{volume}{192}},
  \bibinfo{pages}{912--914}, \doiprefix\url{10.1126/science.192.4242.912}
  (\bibinfo{year}{1976}).

\bibitem{Theraulaz1998}
\bibinfo{author}{Theraulaz, G.}, \bibinfo{author}{Bonabeau, E.} \&
  \bibinfo{author}{Deneubourg, J.-L.}
\newblock \bibinfo{journal}{\bibinfo{title}{The origin of nest complexity in
  social insects}}.
\newblock {\emph{\JournalTitle{Complexity}}} \textbf{\bibinfo{volume}{3}},
  \bibinfo{pages}{15--25},
  \doiprefix\url{10.1002/(sici)1099-0526(199807/08)3:6<15::aid-cplx3>3.0.co;2-v}
  (\bibinfo{year}{1998}).

\bibitem{Czaczkes2015}
\bibinfo{author}{Czaczkes, T.~J.}, \bibinfo{author}{Grüter, C.} \&
  \bibinfo{author}{Ratnieks, F.~L.}
\newblock \bibinfo{journal}{\bibinfo{title}{Trail pheromones: An integrative
  view of their role in social insect colony organization}}.
\newblock {\emph{\JournalTitle{Annual Review of Entomology}}}
  \textbf{\bibinfo{volume}{60}}, \bibinfo{pages}{581--599},
  \doiprefix\url{10.1146/annurev-ento-010814-020627} (\bibinfo{year}{2015}).

\bibitem{Angilletta2007}
\bibinfo{author}{Angilletta, M.~J.}, \bibinfo{author}{II, T. C.~R.},
  \bibinfo{author}{Wilson, R.~S.}, \bibinfo{author}{Niehaus, A.~C.} \&
  \bibinfo{author}{Ribeiro, P.~L.}
\newblock \bibinfo{journal}{\bibinfo{title}{The fast and the fractalous: speed
  and tortuosity trade off in running ants}}.
\newblock {\emph{\JournalTitle{Functional Ecology}}}
  \textbf{\bibinfo{volume}{0}}, \bibinfo{pages}{071027215958004--???},
  \doiprefix\url{10.1111/j.1365-2435.2007.01348.x} (\bibinfo{year}{2007}).

\bibitem{regnier1968insect}
\bibinfo{author}{Regnier, F.~E.} \& \bibinfo{author}{Law, J.~H.}
\newblock \bibinfo{journal}{\bibinfo{title}{Insect pheromones}}.
\newblock {\emph{\JournalTitle{Journal of Lipid Research}}}
  \textbf{\bibinfo{volume}{9}}, \bibinfo{pages}{541--551}
  (\bibinfo{year}{1968}).

\bibitem{Dorigo06antcolony}
\bibinfo{author}{Dorigo, M.}, \bibinfo{author}{Birattari, M.} \&
  \bibinfo{author}{Stützle, T.}
\newblock \bibinfo{journal}{\bibinfo{title}{Ant colony optimization –
  artificial ants as a computational intelligence technique}}.
\newblock {\emph{\JournalTitle{IEEE Comput. Intell. Mag}}}
  \textbf{\bibinfo{volume}{1}}, \bibinfo{pages}{28--39} (\bibinfo{year}{2006}).

\bibitem{Jeanson2005}
\bibinfo{author}{Jeanson, R.} \emph{et~al.}
\newblock \bibinfo{journal}{\bibinfo{title}{Self-organized aggregation in
  cockroaches}}.
\newblock {\emph{\JournalTitle{Animal Behaviour}}}
  \textbf{\bibinfo{volume}{69}}, \bibinfo{pages}{169--180},
  \doiprefix\url{10.1016/j.anbehav.2004.02.009} (\bibinfo{year}{2005}).

\bibitem{Bonabeau1997}
\bibinfo{author}{Bonabeau, E.}, \bibinfo{author}{Theraulaz, G.},
  \bibinfo{author}{Deneubourg, J.-L.}, \bibinfo{author}{Aron, S.} \&
  \bibinfo{author}{Camazine, S.}
\newblock \bibinfo{journal}{\bibinfo{title}{Self-organization in social
  insects}}.
\newblock {\emph{\JournalTitle{Trends in Ecology {\&} Evolution}}}
  \textbf{\bibinfo{volume}{12}}, \bibinfo{pages}{188--193},
  \doiprefix\url{10.1016/s0169-5347(97)01048-3} (\bibinfo{year}{1997}).

\bibitem{Bartholdi1993}
\bibinfo{author}{Bartholdi, J.~J.}, \bibinfo{author}{Seeley, T.~D.},
  \bibinfo{author}{Tovey, C.~A.} \& \bibinfo{author}{Vate, J.~H.}
\newblock \bibinfo{journal}{\bibinfo{title}{The pattern and effectiveness of
  forager allocation among flower patches by honey bee colonies}}.
\newblock {\emph{\JournalTitle{Journal of Theoretical Biology}}}
  \textbf{\bibinfo{volume}{160}}, \bibinfo{pages}{23--40},
  \doiprefix\url{10.1006/jtbi.1993.1002} (\bibinfo{year}{1993}).

\bibitem{Beckers1990}
\bibinfo{author}{Beckers, R.}, \bibinfo{author}{Deneubourg, J.~L.},
  \bibinfo{author}{Goss, S.} \& \bibinfo{author}{Pasteels, J.~M.}
\newblock \bibinfo{journal}{\bibinfo{title}{Collective decision making through
  food recruitment}}.
\newblock {\emph{\JournalTitle{Insectes Sociaux}}}
  \textbf{\bibinfo{volume}{37}}, \bibinfo{pages}{258--267},
  \doiprefix\url{10.1007/bf02224053} (\bibinfo{year}{1990}).

\bibitem{Beckers1993}
\bibinfo{author}{Beckers, R.}, \bibinfo{author}{Deneubourg, J.~L.} \&
  \bibinfo{author}{Goss, S.}
\newblock \bibinfo{journal}{\bibinfo{title}{Modulation of trail laying in the
  antlasius niger (hymenoptera: Formicidae) and its role in the collective
  selection of a food source}}.
\newblock {\emph{\JournalTitle{Journal of Insect Behavior}}}
  \textbf{\bibinfo{volume}{6}}, \bibinfo{pages}{751--759},
  \doiprefix\url{10.1007/BF01201674} (\bibinfo{year}{1993}).

\bibitem{Beekman2001}
\bibinfo{author}{Beekman, M.}, \bibinfo{author}{Sumpter, D. J.~T.} \&
  \bibinfo{author}{Ratnieks, F. L.~W.}
\newblock \bibinfo{journal}{\bibinfo{title}{Phase transition between disordered
  and ordered foraging in pharaoh's ants}}.
\newblock {\emph{\JournalTitle{Proceedings of the National Academy of
  Sciences}}} \textbf{\bibinfo{volume}{98}}, \bibinfo{pages}{9703--9706},
  \doiprefix\url{10.1073/pnas.161285298} (\bibinfo{year}{2001}).

\bibitem{Bonabeau1996}
\bibinfo{author}{Bonabeau, E.}
\newblock \bibinfo{journal}{\bibinfo{title}{Comment on "phase transitions in
  instigated collective decision making"}}.
\newblock {\emph{\JournalTitle{Adaptive Behavior}}}
  \textbf{\bibinfo{volume}{5}}, \bibinfo{pages}{99--105},
  \doiprefix\url{10.1177/105971239600500105} (\bibinfo{year}{1996}).

\bibitem{Deneubourg1990}
\bibinfo{author}{Deneubourg, J.~L.}, \bibinfo{author}{Aron, S.},
  \bibinfo{author}{Goss, S.} \& \bibinfo{author}{Pasteels, J.~M.}
\newblock \bibinfo{journal}{\bibinfo{title}{The self-organizing exploratory
  pattern of the argentine ant}}.
\newblock {\emph{\JournalTitle{Journal of Insect Behavior}}}
  \textbf{\bibinfo{volume}{3}}, \bibinfo{pages}{159--168},
  \doiprefix\url{10.1007/BF01417909} (\bibinfo{year}{1990}).

\bibitem{Sumpter2003}
\bibinfo{author}{Sumpter, D.} \& \bibinfo{author}{Pratt, S.}
\newblock \bibinfo{journal}{\bibinfo{title}{A modelling framework for
  understanding social insect foraging}}.
\newblock {\emph{\JournalTitle{Behavioral Ecology and Sociobiology}}}
  \textbf{\bibinfo{volume}{53}}, \bibinfo{pages}{131--144},
  \doiprefix\url{10.1007/s00265-002-0549-0} (\bibinfo{year}{2003}).

\bibitem{Perna2012}
\bibinfo{author}{Perna, A.} \emph{et~al.}
\newblock \bibinfo{journal}{\bibinfo{title}{Individual rules for trail pattern
  formation in argentine ants (linepithema humile)}}.
\newblock {\emph{\JournalTitle{{PLoS} Computational Biology}}}
  \textbf{\bibinfo{volume}{8}}, \bibinfo{pages}{e1002592},
  \doiprefix\url{10.1371/journal.pcbi.1002592} (\bibinfo{year}{2012}).

\bibitem{Beckers1992}
\bibinfo{author}{Beckers, R.}, \bibinfo{author}{Deneubourg, J.~L.} \&
  \bibinfo{author}{Goss, S.}
\newblock \bibinfo{journal}{\bibinfo{title}{Trail laying behaviour during food
  recruitment in the antlasius niger (l.)}}.
\newblock {\emph{\JournalTitle{Insectes Sociaux}}}
  \textbf{\bibinfo{volume}{39}}, \bibinfo{pages}{59--72},
  \doiprefix\url{10.1007/BF01240531} (\bibinfo{year}{1992}).

\bibitem{Calenbuhr1992}
\bibinfo{author}{Calenbuhr, V.} \& \bibinfo{author}{Deneubourg, J.-L.}
\newblock \bibinfo{journal}{\bibinfo{title}{A model for osmotropotactic
  orientation (i)}}.
\newblock {\emph{\JournalTitle{Journal of Theoretical Biology}}}
  \textbf{\bibinfo{volume}{158}}, \bibinfo{pages}{359--393},
  \doiprefix\url{10.1016/s0022-5193(05)80738-6} (\bibinfo{year}{1992}).

\bibitem{dias2012intersecting}
\bibinfo{author}{Dias, C.}, \bibinfo{author}{Sarvi, M.} \&
  \bibinfo{author}{Shiwakoti, N.}
\newblock \bibinfo{title}{Intersecting and merging pedestrian crowd flows under
  panic conditions: insights from biological entities}.
\newblock In \emph{\bibinfo{booktitle}{35th Australasian Transport Research
  Forum (ATRF)}}, \bibinfo{pages}{13--20} (\bibinfo{organization}{Perth,
  Western Australia Australia}, \bibinfo{year}{2012}).

\bibitem{Hu2003}
\bibinfo{author}{Hu, X.~P.}, \bibinfo{author}{Appel, A.~G.} \&
  \bibinfo{author}{Traniello, J. F.~A.}
\newblock \bibinfo{journal}{\bibinfo{title}{Behavioral response of two
  subterranean termites (isoptera: Rhinotermitidae) to vibrational stimuli}}.
\newblock {\emph{\JournalTitle{Journal of Insect Behavior}}}
  \textbf{\bibinfo{volume}{16}}, \bibinfo{pages}{703--715},
  \doiprefix\url{10.1023/B:JOIR.0000007705.50488.57} (\bibinfo{year}{2003}).

\bibitem{Schwinghammer2006}
\bibinfo{author}{Schwinghammer, M.~A.} \& \bibinfo{author}{Houseman, R.~M.}
\newblock \bibinfo{journal}{\bibinfo{title}{Response of ireticulitermes
  flavipes/i (isoptera: Rhinotermitidae) to disturbance in laboratory arenas at
  different temperatures and soldier proportions}}.
\newblock {\emph{\JournalTitle{Journal of Economic Entomology}}}
  \textbf{\bibinfo{volume}{99}}, \bibinfo{pages}{462--468},
  \doiprefix\url{10.1603/0022-0493-99.2.462} (\bibinfo{year}{2006}).

\bibitem{Gautam2011}
\bibinfo{author}{Gautam, B.~K.} \& \bibinfo{author}{Henderson, G.}
\newblock \bibinfo{journal}{\bibinfo{title}{Escape behavior of the formosan
  subterranean termite (isoptera: Rhinotermitidae) in response to
  disturbance}}.
\newblock {\emph{\JournalTitle{Journal of Insect Behavior}}}
  \textbf{\bibinfo{volume}{25}}, \bibinfo{pages}{70--79},
  \doiprefix\url{10.1007/s10905-011-9278-4} (\bibinfo{year}{2011}).

\bibitem{Wang2015}
\bibinfo{author}{Wang, C.}, \bibinfo{author}{Henderson, G.},
  \bibinfo{author}{Gautam, B.~K.}, \bibinfo{author}{Chen, J.} \&
  \bibinfo{author}{Bhatta, D.}
\newblock \bibinfo{journal}{\bibinfo{title}{Panic escape polyethism in worker
  and {soldierCoptotermes} formosanus(isoptera: Rhinotermitidae)}}.
\newblock {\emph{\JournalTitle{Insect Science}}} \textbf{\bibinfo{volume}{23}},
  \bibinfo{pages}{305--312}, \doiprefix\url{10.1111/1744-7917.12206}
  (\bibinfo{year}{2015}).

\bibitem{Nakamura1972}
\bibinfo{author}{Nakamura, E.~L.}
\newblock \emph{\bibinfo{title}{Development and Uses of Facilities for Studying
  Tuna Behavior}}, \bibinfo{pages}{245--277} (\bibinfo{publisher}{Springer},
  \bibinfo{year}{1972}).

\bibitem{Tejera2016}
\bibinfo{author}{Tejera, F.}, \bibinfo{author}{Reyes, A.} \&
  \bibinfo{author}{Altshuler, E.}
\newblock \bibinfo{journal}{\bibinfo{title}{Uninformed sacrifice: Evidence
  against long-range alarm transmission in foraging ants exposed to localized
  abduction}}.
\newblock {\emph{\JournalTitle{The European Physical Journal Special Topics}}}
  \textbf{\bibinfo{volume}{225}}, \bibinfo{pages}{663--668},
  \doiprefix\url{10.1140/epjst/e2015-50325-8} (\bibinfo{year}{2016}).

\end{thebibliography}
\end{document}